# Electron-Neutrino Scattering a Strong Aspirant for Precision Measurements

Abrar Ahmad
*Department of Physics, COMSATS University Islamabad, Pakistan. and Islamabad, Pakistan. E-mail:smart5733@gmail.com*

Shakeel Mahmood
*Department of Physics, Air University, PAF Complex, E-9, Islamabad, Pakistan. and E-mail:shakeel.mahmood@mail.au.edu.pk*

Farida Tahir
*Department of Physics, COMSATS University, Islamabad, Pakistan. and E-mail: farida_tahir@comsats.edu.pk*

Imama Ijaz
*Department of Physics, Air University, PAF Complex, E-9, Islamabad, Pakistan. and E-mail: 200024@students.au.edu.pk*

Wasi Uz Zaman
*Department of Physics, Air University, PAF Complex, E-9, Islamabad, Pakistan. and E-mail: 190198@students.au.edu.pk*

This work focuses on the study $\nu_\alpha\, e \to \nu_\beta\, e$ scattering in the frame work of physics beyond the standard model (SM) called new physics (NP). Both Model Independent (MI) and Model Dependent (MD) ways are used to constrain NP. R-parity violating Supersymmetry ($\not{R}_p$ SUSY) Model is used to perform MD analysis, where the scattering cross-section is influenced by new s-bosons. For MI way non-standard neutrino intections (NSI) are used where there is no need of introducing any new particle. LSND and LAMPF-E225 data is used to identify the physically allowed and forbidden regions for nonunivarsal NSI parameters $\acute{\varepsilon}^{eP}_{ee}$ $(P = R, R)$ and nonunivarsal SUSY parameters $\sum_j |\lambda_{1j1}|$. Similarly, CHARM-II, BNL-COL and BNL-E734 experimental is used to explore allowed and forbidden regions for $\acute{\varepsilon}^{eP}_{\mu\mu}$ , $\sum_k |\lambda_{21k}|$ and $\sum_j |\lambda_{2j1}|$ and limits are established. Furthermore, we establish a relationship between MI and MD coupling parameters.

## 1. INTRODUCTION

Current precision demands a complete understanding of model parameters in addition to their mapping with respect to experimental results. To achieve this, all the possible theoretical information for the interpretation of data is incorporated [1–4]. The compatibility between experimental data and theoretical model is only valid within the limits of experimental uncertainties. There are discrepancies between the predictions of SM and the observations made by Liquid Scintillator Neutrino Detector (LSND) [2], Los Alamos Meson Physics Facility (LAMPF)[3]. CHARM-II[48], BNL-COL[56] and BNL-E734 [49] experiments. Standard Model (SM) treats neutrinos as massless particles, that fortifies the fact that not only lepton number and flavor will be conserved, but also $W^{\pm}, Z$ weak gauge bosons universally couple with all three families of leptons [5, 6]. Contrary to that, Sudbury Neutrino Observatory (SNO) observed the flavor changing for $^8B$ solar neutrinos, during their flight from the center of the Sun to the Earth [7]. Super-Kamiokande (SK) observed two-flavor $\nu_\mu \longleftrightarrow \nu_\tau$ atmospheric neutrino oscillations [8–11]. Kamioka Liquid Scintillator Anti neutrino Detector (KamLAND), KEK to Kamioka (K2K) and Main Injector Neutrino Oscillation Search (MINOS) have measured different neutrino oscillation parameters with extraordinary precision [12]. In solar and atmospheric neutrino experiments, lepton flavour violation has been observed [4]. These evidences of neutrino oscillations implies finite neutrinos masses and mixing [13].

In the light of these facts, it is a clearly established fact that there was something wrong with the naive SM description of neutrino properties at the very fundamental level. Therefore, SM either needs to be modified or extended to incorporate new interactions (NI). Such a model can be possible beyond the SM that distinguish among the three generations of neutrinos, and violate neutral current universality via the direct exchange of new particles between the neutrinos and matter particles [14–34]. So, many extensions (new physics models (NPM)) of the SM with non-minimal Higgs bosons are available in the form of see-saw models, Left- Right symmetric, GUTs, 331 models and Minimal Super Symmetric Standard Model (MSSM) [35–41] capable to accommodate neutrino masses in the natural way and predicting NI at the same time.

In MSSM, operators that carry the same Baryon number ($B$) and Lepton number ($L$) of the SM but different spin and mass, also violate $B$ and $L$ conservation, which have dangerous impact on the low energy phenomenology. In particular, it means that matter is unstable due to the proton decay. To cope with this problem an ad hoc discrete symmetry namely R-parity ($R_p = (-1)^{3B+L+2S}$) introduced, with $B$, $L$ and intrinsic spin ($S$) of the particle. Under this condition $L$ and $B$ number violating processes are prohibited thus preventing proton from decay [42]. As R-parity is ad hoc symmetry, therefore in order to extend the SM and study Beyond the SM, the possibility of B and/or L-violation must be encountered. As an alternate solution of rapid proton decay problem, relaxed $R_p$ conservation by assuming $B$-conservation but $L$-violation. This alleviation allowed the Yukawa couplings $\lambda_{ijk}$ ($LLE$), $\lambda'_{ijk}$ ($LQD$), their product $\left(\lambda_{ijk}\, \lambda_{\alpha\beta\gamma}\ ,\ \lambda'_{ijk}\, \lambda'_{\alpha\beta\gamma}\right)$ and combinations $\left(\lambda_{ijk}\ \lambda'_{ijk}\right)$ forbidden due to simultaneous presence of $\lambda'_{ijk}$ $\lambda''_{ijk}$ ($LQD$ and $UDD$) and $\lambda_{ijk}$ $\lambda''_{ijk}$ ($LLE$ and $UDD$) in most general $\not{R}_p$ superpotential Eq.(1).

$$W_{\not{R}_p} = \frac{1}{2}\lambda_{[ij]k}\, L_i\, L_i\, e^c_k + \lambda'_{ijk}\, L_i\, Q_j\, d^c_k + \frac{1}{2}\lambda''_{i[jk]}\, u^c_i\, d^c_j\, d^c_k \tag{1}$$

where $i, j, k = (1, 2, 3)$ represents three generation of quarks and lepton chiral super fields, $L_i$ , $Q_i$ , $d^c_i$ ,$d^c_i$ and $e^c_i$ . Superscript "$c$" denote the charge conjugate spinor. The first two terms are responsible for the lepton number violation and last term is responsible for baryon number violation. The $SU(2)_L$ gauge invariance demands the couplings $\lambda_{ijk}$ to be antisymmetric in the first two indices, whereas $SU(3)_C$ gauge invariance requires the couplings $\lambda''_{ijk}$ to be antisymmetric in the last two indices. These conditions reduce the number of $\not{R}_p$ couplings to 45 ($9\lambda_{ijk}$ , $27\lambda'_{ijk}$ ,and $9\lambda''_{ijk}$ ) [43–47].

The $/R_p$ SUSY model allows flavor changing non-standard interaction (FCNSI), flavor diagonal non-universial non-standard interaction (FDNSI) at tree level and distinguish the three flavors of neutrino in interactions via direct exchanging virtual SUSY propagator. This may explain data in a better way and enrich our understanding about neutrinos. We study such effects both in MI and MD interactions by using reaction $\nu_\alpha\, e \to \nu_\beta\, e$ ($\alpha = \beta = e, \mu$). We have selected $\nu_e\, e \to \nu_e\, e$ reaction for two reasons:

(1) It is a reliable source for the precision measurements and

(2) It is the only reaction (in the history of particle physics), which can proceed through both charge and neutral current interactions and confirm (support) the gauge structure of the SM.

This paper is organized as follow: In Sec-II we perform an analysis of total elastic scattering cross-section for $\nu_e\, e$ and $\nu_\mu\, e$ in the context of SM and discuss a mismatch between experiments and SM prediction. In Sec. III, we perform a detail analysis of $\nu_\alpha\, e \to \nu_\beta\, e$ in MI way and investigate the role of NSI. We also discuss the physically allowed and forbidden regions for NSI parameters and discuss the sensitivity of these parameters. In Sec. IV, we calculate MD $\nu_e\, e$ and $\nu_\mu\, e$ elastic scattering cross-section by using $\not{R}_p$ SUSY model and obtain physically allowed and forbidden regions on Non Universal Flavor Diagonal (NUFD) $\not{R}_p$ SUSY couplings parameters. Finally we give results and discuusion followed by summary.

## 2. STANDARD MODEL CROSS-SECTION

Neutrino-Electron scattering( $\nu_\alpha\, e \to \nu_\beta\, e$) are the simplest of purely leptonic weak processes, which are favorable for the precision test measurement and search for physics beyond SM. These include both charge current (CC) and neutral current (NC). Both CC and NC contribute in case of electron neutrino i.e. $\alpha = \beta = e$, and only NC contribution in case of muon and tau neutrinos i.e. $\alpha = \beta = \mu, \tau$ within SM[57, 58]. The Feynman diagrams of these processes are shown in Fig.(1)

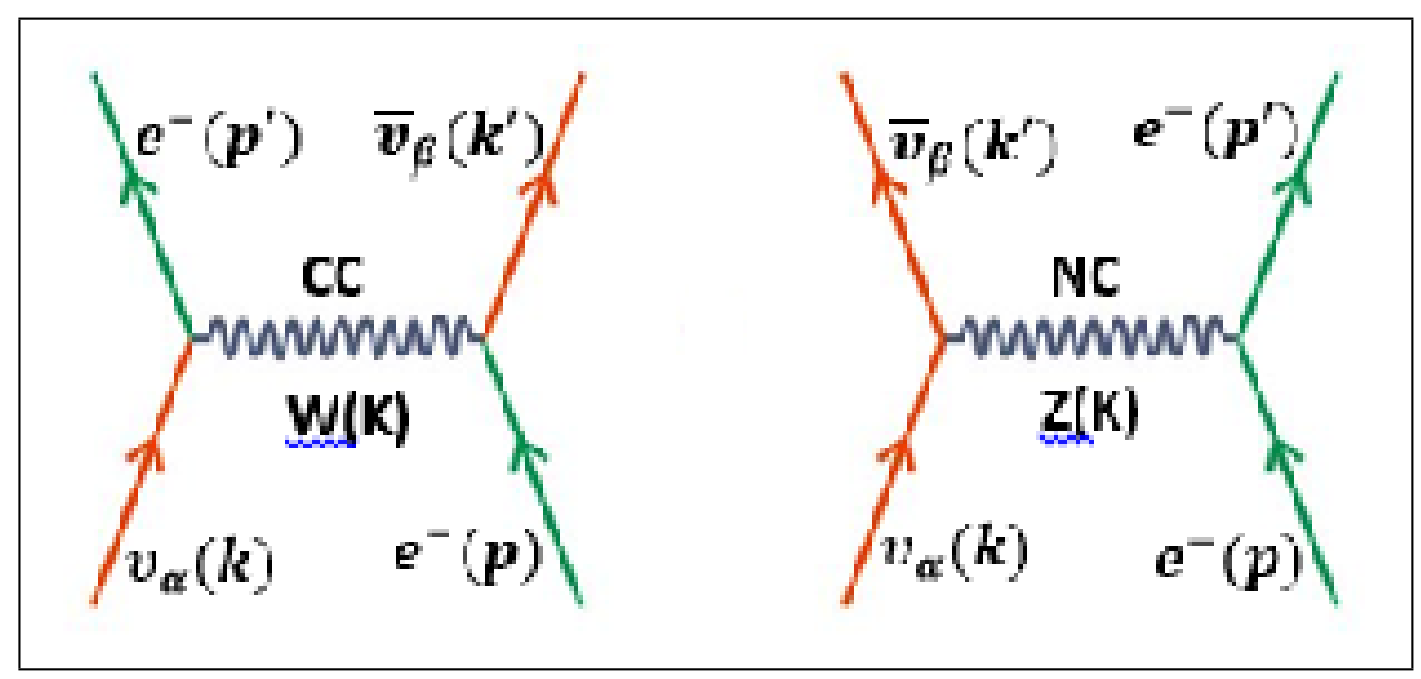

| Experiment | $\sigma(\nu_e)/\mathbf{E}_\nu \times \mathbf{10}^{-45} \times \mathbf{cm}^2\mathbf{MeV}^{-1}$ | SM Contribution | Uncertainty |
|---|---|---|---|
| LSND [2] | $10.1 \pm 1.1 \pm 1.0$ | ∼94.3% | $\sim -5.7\%$ |
| LAMPF E225[3] | $10.0 \pm 1.5 \pm 0.9$ | ∼95.2% | $\sim -4.8\%$ |
| | $\sigma(\nu_\mu)/\mathbf{E}_\nu \times \mathbf{10}^{-45} \times \mathbf{cm}^2\mathbf{MeV}^{-1}$ | | |
| CHARM-II [48] | $1.53 \pm 0.04 \pm 0.12$ | ∼101.6% | $\sim +1.6\%$ |
| BNL-COL[56] | $1.67 \pm 0.44$ | ∼93.1% | $\sim -6.9\%$ |
| BNL-E734 [49] | $1.8 \pm 0.2 \pm 0.25$ | ∼86.3% | $\sim -13.7\%$ |

TABLE I: Measured cross-section for different experiments and SM contribution and its deviation from central experimental values is summarized here in tabular form

Fig. (1): Tree-level diagrams that contribute to the $\nu_\alpha\, e$ scattering in the standard model [13].

In the light of SM, the effective Lagrangians of CC and NC are [59];

$$^{CC}L_{eff} = -2\sqrt{2}G_F\, [\bar{\nu}_e\, (k')\gamma_a\, P_L\, \nu_e\, (k)][\bar{e}(p')\gamma^\alpha P_L\, e(p)] \tag{2}$$

$$^{NC}L_{eff} = -\sqrt{2}G_F\, [\bar{\nu}_\alpha\, (k')\gamma_\mu\, P_L\, \nu_\alpha\, (k)]\, [(\bar{e}(p')\gamma^\mu(g_L\, P_L + g_R\, P_R\,)e(p)] \tag{3}$$

Where $G_F$ is Fermi coupling constant, $P_L \equiv \left(\frac{1-\gamma^5}{2}\right)$ and $P_R \equiv \left(\frac{1+\gamma^5}{2}\right)$ are left and right handed chiral operators respectively [59], $g_L = g_V + g_A$ ; $g_R = g_V + g_A$ ; $g_V$ and $g_A$ are vector and axial-vector coupling constants respectively. In SM, $g_V = -\frac{1}{2} + 2\sin^2\theta_w$, $g_A = -\frac{1}{2}$ and $\sin^2\theta_w = 0.23$ is the weak mixing angle [2]. The total Lagrangian of CC and NC for $\nu_e\, e \to \nu_e\, e$ within SM is

$$\begin{aligned} L_{SM} &= ^{CC} L_{eff} + ^{NC} L_{eff} \\ &= -\sqrt{2}G_F\, [\bar{\nu}_e(k')\gamma_\mu\, P_L\, \nu_e\, (k)]\, [\bar{e}(p')\gamma^\mu(A_L\, P_L + A_R\, P_R\,)e(p)] \end{aligned} \tag{4}$$

Where $A_L \equiv 2 + g_L$ and $A_R \equiv g_R$ . Using Eq.(4) we get the total elastic cross-section for $\nu_e\, e$ as

$$\sigma_{SM}^{v_e} = \sigma_o \left[(g_L + 2)^2 + \frac{g_R^2}{3}\right] \tag{5}$$

and for $\nu_\alpha\, e \to \nu_\alpha\, e$ $(\alpha = \mu)$ we get total cross-section by using Eq.(3)

$$\sigma_{SM}^{v_\mu} = \sigma_o \left[g_L^2 + \frac{g_R^2}{3}\right] \tag{6}$$

where $\sigma_o \equiv \frac{G_F^2 m_e E_\nu}{2\pi} = 4.31 \times 10^{-45} E_v cm^2/MeV$ ($E_\nu$ is the incident energy of neutrinos). We get SM cross-section for $\nu_e\, e \to \nu_e\, e$ and $\nu_\mu\, e \to \nu_\mu\, e$ processes are

$$\sigma^{v_e} = 9.52 \times 10^{-45} \left(cm^2/MeV\right) \times E_\nu \tag{7}$$

$$\sigma^{v_\mu} = 1.55 \times 10^{-45} \left(cm^2/MeV\right) \times E_\nu. \tag{8}$$

The measured cross-section related to different experiments are shown in Table-I. The difference in measured cross-section from the SM at different energies and sigma levels is shown in Figs.(2-6). This deviation is guiding us to incorporate non-standard effects in our theory. The uncertainity in theoretical value is coming from $G_F$ , $m_e$ , $g_L$ and $g_R$ values.

## 3. MODEL INDEPENDENT ANALYSIS OF $\nu_\alpha\, e \to \nu_\beta\, e$

NSI play an important role in the precision of measurements. The discrepancy between the theoretical and experimental values of $\nu_e\, e$ and $\nu_\mu\, e$ cross-sections can be treated as NSI which is parametrised in terms of unknown coupling parameters ($\varepsilon_{\alpha\beta}^{eL}$ , $\varepsilon_{\alpha\beta}^{eR}$ ), which describe the strength of NSI. The Feynman diagram of NSI for $\nu_\alpha\, e \to \nu_\beta\, e$ is shown in Fig.(2).

| CL level | Limits on NSI parameters $\acute{\varepsilon}^{eL}_{ee}$ and $\acute{\varepsilon}^{eR}_{ee}$ | | | |
|---|---|---|---|---|
| | LSND | | LAMPF E225 | |
| 68% | -1.54$\leq \acute{\varepsilon}^{eL}_{ee} \leq$-1.42<br>-0.04 $\leq \acute{\varepsilon}^{eR}_{ee} \leq$0.08 | -0.87$\leq \acute{\varepsilon}^{eR}_{ee} \leq$0.41 | -1.55$\leq \acute{\varepsilon}^{eL}_{ee} \leq$-1.41<br>-0.05 $\leq \acute{\varepsilon}^{eR}_{ee} \leq$0.08 | -0.89$\leq \acute{\varepsilon}^{eR}_{ee} \leq$0.43 |
| 90% | -1.57$\leq \acute{\varepsilon}^{eL}_{ee} \leq$-1.38<br>-0.08 $\leq \acute{\varepsilon}^{eR}_{ee} \leq$0.11 | -0.99$\leq \acute{\varepsilon}^{eR}_{ee} \leq$0.53 | -1.58$\leq \acute{\varepsilon}^{eL}_{ee} \leq$-1.36<br>-0.10 $\leq \acute{\varepsilon}^{eR}_{ee} \leq$0.12 | -1.03$\leq \acute{\varepsilon}^{eR}_{ee} \leq$0.57 |
| 95% | -1.59$\leq \acute{\varepsilon}^{eL}_{ee} \leq$-1.36<br>-0.10 $\leq \acute{\varepsilon}^{eR}_{ee} \leq$0.13 | -1.04$\leq \acute{\varepsilon}^{eR}_{ee} \leq$0.58 | -1.60$\leq \acute{\varepsilon}^{eL}_{ee} \leq$-1.33<br>-0.13 $\leq \acute{\varepsilon}^{eR}_{ee} \leq$0.14 | -1.09$\leq \acute{\varepsilon}^{eR}_{ee} \leq$0.63 |

TABLE II: NSI coupling parameter limits at different CL levels

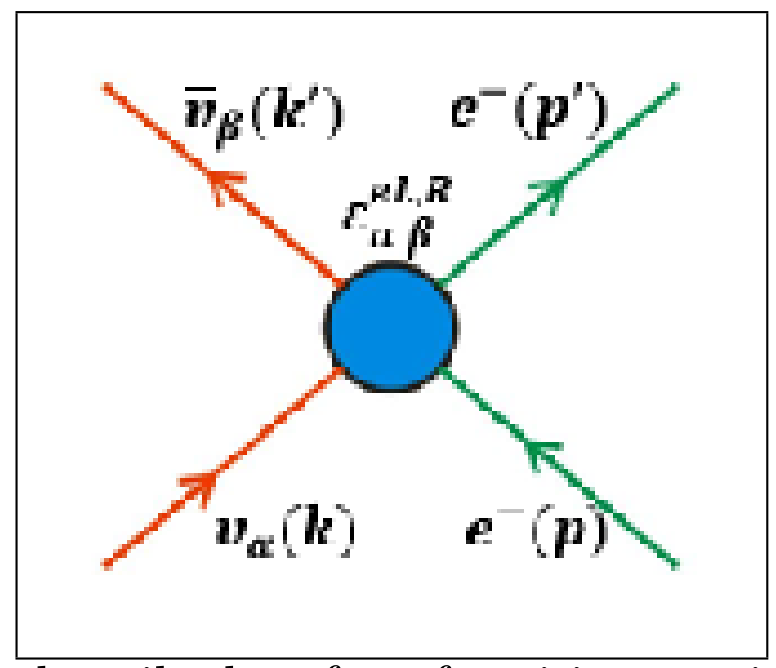

Fig. (2): NSI of neutrinos described as four-fermi interaction with new couplings [13]

The Lagrangian of Non Standard neutral current Interactions can be written as;

$$L_{(NSI)} = -\sqrt{2}G_F \left[\bar{\nu}_\beta \ (k')\gamma_\mu \ P_L \nu_\alpha \ (k)\right] \left[\bar{e}(p')\gamma^\mu (\varepsilon^{eL}_{\alpha\beta} \ P_L + \varepsilon^{eR}_{\alpha\beta} \ P_R) e(p)\right] \tag{9}$$

where $\alpha$ and $\beta$ represent the lepton flavor ($e$, $\mu$ *or* $\tau$) $\alpha = \beta$ represents Non Universal (NU) Flavor-conserving NSI and $\alpha \neq \beta$ for Flavor-Changing (FC) NSI [12]. SM contribution to the cross section of $\nu_e$ and $\nu_\mu$ is taken as the most reliable contribution and therefore, NSI is considered as a sub-leading effect. So, NSI are added as corrections in Eqs. (11) and (12).

We are intrested in elastic scattering therfore neglecting Flavor-Changing NSI contribution. The elastic scattering cross-section for $\nu_e e$ and $\nu_\mu e$ with contribution of NU flavor conserving NSI are given by;

$$\sigma^{v_e} = 4\sigma_o \ \left[\left(1 + \frac{g_L}{2} + \acute{\varepsilon}^{eL}_{ee}\right)^2 + \frac{1}{3}\left(\frac{g_R}{2} + \acute{\varepsilon}^{eR}_{ee}\right)^2\right] \tag{10}$$

$$\sigma^{v_\mu} = 4\sigma_o \ \left[\left(\frac{g_L}{2} + \acute{\varepsilon}^{eL}_{\mu\mu}\right)^2 + \frac{1}{3}\left(\frac{g_R}{2} + \acute{\varepsilon}^{eR}_{\mu\mu}\right)^2\right] \tag{11}$$

where $\acute{\varepsilon}^{eL,R}_{\alpha\beta} \equiv \frac{\varepsilon^{eL,R}_{\alpha\beta}}{2}$ with ($\alpha = \beta = e$ *or* $\mu$).

The dependency of $\acute{\varepsilon}^{eR}_{ee}$ to the $\acute{\varepsilon}^{eL}_{ee}$ can be obtain by using the LSND and LAMPF E225 $\nu_e \ e \rightarrow \nu_e \ e$ experimental value from Table-I in Eq.(10) as

$$\acute{\varepsilon}^{eR}_{ee} = \pm\sqrt{\frac{3\sigma^{v_e}_{\text{exp}}}{4\sigma_o} - 3\left(1 + \frac{g_L}{2} + \acute{\varepsilon}^{eL}_{ee}\right)^2} - \frac{g_R}{2} \tag{12}$$

using Eq.(12), we obtaine allowed regions at 90% confidence level (CL) which shows sensitivity of $\acute{\varepsilon}^{eR}_{ee}$ on $\acute{\varepsilon}^{eL}_{ee}$.

The limits on flavour conserving NSI parameters calculated at 68%CL, 90%CL and 95%CL by using single coupling dominance hypothesis mentioned in Table-II. These limits at 90% from LSND experiment compatible with reference [4].

In a similar way we extract dependency of $\acute{\varepsilon}^{eR}_{\mu\mu}$ to the $\acute{\varepsilon}^{eL}_{\mu\mu}$ by using CHARM-II, BNL-COL and BNL-E734 experiments from Table-1 in Eq.(11 ) is reversed as

$$\acute{\varepsilon}^{eR}_{\mu\mu} = \pm\sqrt{\frac{3\sigma^{v_\mu}_{\text{exp}}}{4\sigma_o} - 3\left(\frac{g_L}{2} + \acute{\varepsilon}^{eL}_{\mu\mu}\right)^2} - \frac{g_R}{2} \tag{13}$$

| CL level | Limits on NSI parameters $\acute{\varepsilon}^{eL}_{\mu\mu}$ and $\acute{\varepsilon}^{eR}_{\mu\mu}$ | | | | | |
|---|---|---|---|---|---|---|
| | CHRAM-II | | BNL-COL | | BNL-E734 | |
| 68% | -0.01≤ $\acute{\varepsilon}^{eL}_{\mu\mu}$ ≤0.02<br>0.52≤ $\acute{\varepsilon}^{eR}_{\mu\mu}$ ≤0.55 | -0.50≤ $\acute{\varepsilon}^{eR}_{\mu\mu}$ ≤-0.40<br>-0.06≤ $\acute{\varepsilon}^{eR}_{\mu\mu}$ ≤0.04 | -0.05≤ $\acute{\varepsilon}^{eL}_{\mu\mu}$ ≤0.04<br>0.50≤ $\acute{\varepsilon}^{eR}_{\mu\mu}$ ≤0.59 | -0.62≤ $\acute{\varepsilon}^{eR}_{\mu\mu}$ ≤0.16 | -0.06≤ $\acute{\varepsilon}^{eL}_{\mu\mu}$ ≤0.01<br>0.53≤ $\acute{\varepsilon}^{eR}_{\mu\mu}$ ≤0.59 | -0.62≤ $\acute{\varepsilon}^{eR}_{\mu\mu}$ ≤-0.44<br>-0.03≤ $\acute{\varepsilon}^{eR}_{\mu\mu}$ ≤0.16 |
| 90% | -0.02≤ $\acute{\varepsilon}^{eL}_{\mu\mu}$ ≤0.03<br>0.51≤ $\acute{\varepsilon}^{eR}_{\mu\mu}$ ≤0.56 | -0.53≤ $\acute{\varepsilon}^{eR}_{\mu\mu}$ ≤-0.35<br>-0.11≤ $\acute{\varepsilon}^{eR}_{\mu\mu}$ ≤0.06 | -0.08≤ $\acute{\varepsilon}^{eL}_{\mu\mu}$ ≤0.08<br>0.46≤ $\acute{\varepsilon}^{eR}_{\mu\mu}$ ≤0.62 | -0.68≤ $\acute{\varepsilon}^{eR}_{\mu\mu}$ ≤0.22 | -0.07≤ $\acute{\varepsilon}^{eL}_{\mu\mu}$ ≤0.03<br>0.50≤ $\acute{\varepsilon}^{eR}_{\mu\mu}$ ≤0.61 | -0.67≤ $\acute{\varepsilon}^{eR}_{\mu\mu}$ ≤-0.31<br>-0.15≤ $\acute{\varepsilon}^{eR}_{\mu\mu}$ ≤0.20 |
| 95% | -0.02≤ $\acute{\varepsilon}^{eL}_{\mu\mu}$ ≤0.03<br>0.50≤ $\acute{\varepsilon}^{eR}_{\mu\mu}$ ≤0.56 | -0.54≤ $\acute{\varepsilon}^{eR}_{\mu\mu}$ ≤-0.32<br>-0.14≤ $\acute{\varepsilon}^{eR}_{\mu\mu}$ ≤0.07 | -0.09≤ $\acute{\varepsilon}^{eL}_{\mu\mu}$ ≤0.10<br>0.44≤ $\acute{\varepsilon}^{eR}_{\mu\mu}$ ≤0.63 | -0.71≤ $\acute{\varepsilon}^{eR}_{\mu\mu}$ ≤0.24 | -0.08≤ $\acute{\varepsilon}^{eL}_{\mu\mu}$ ≤0.04<br>0.49≤ $\acute{\varepsilon}^{eR}_{\mu\mu}$ ≤0.62 | -0.69≤ $\acute{\varepsilon}^{eR}_{\mu\mu}$ ≤0.22 |

TABLE III: NSI coupling parameter limits at different CL levels

By using Eq.(13), we obtained the graph of allowed regions at 90% CL which shows sensitivity of $\acute{\varepsilon}^{eR}_{\mu\mu}$ on $\acute{\varepsilon}^{eL}_{\mu\mu}$ .

The limits on flavour conserving NSI parameters $\acute{\varepsilon}^{eL}_{\mu\mu}$ and $\acute{\varepsilon}^{eR}_{\mu\mu}$ calculated at 68%CL, 90%CL and 95%CL by using single coupling dominance hypothesis mentioned in Table-III.

## 4. MODEL DEPENDENT ANALYSIS OF $\nu_\alpha e \to \nu_\beta e$

MD analysis is performed in $\not{R}_p$ SUSY Model. The particle content of the Minimal Supersymmetric Standard Model (MSSM) is shown in Eq.(1) involved only the tri-linear couplings in terms of the component fields (with the s-fermion fields characterized by *tilda* sign), the trilinear terms lead to interaction of the form

$$L_{(\not{R}_p)} = \lambda_{ijk} \left[\tilde{\nu}_{iL}\, \bar{e}_{kR}\, e_{jL} + \tilde{e}_{jL}\, \bar{e}_{kR}\, \upsilon_{iL} + \tilde{e}^*_{kR}\, (\bar{\upsilon}_{iL})^c e_{jL}\right] \tag{14}$$

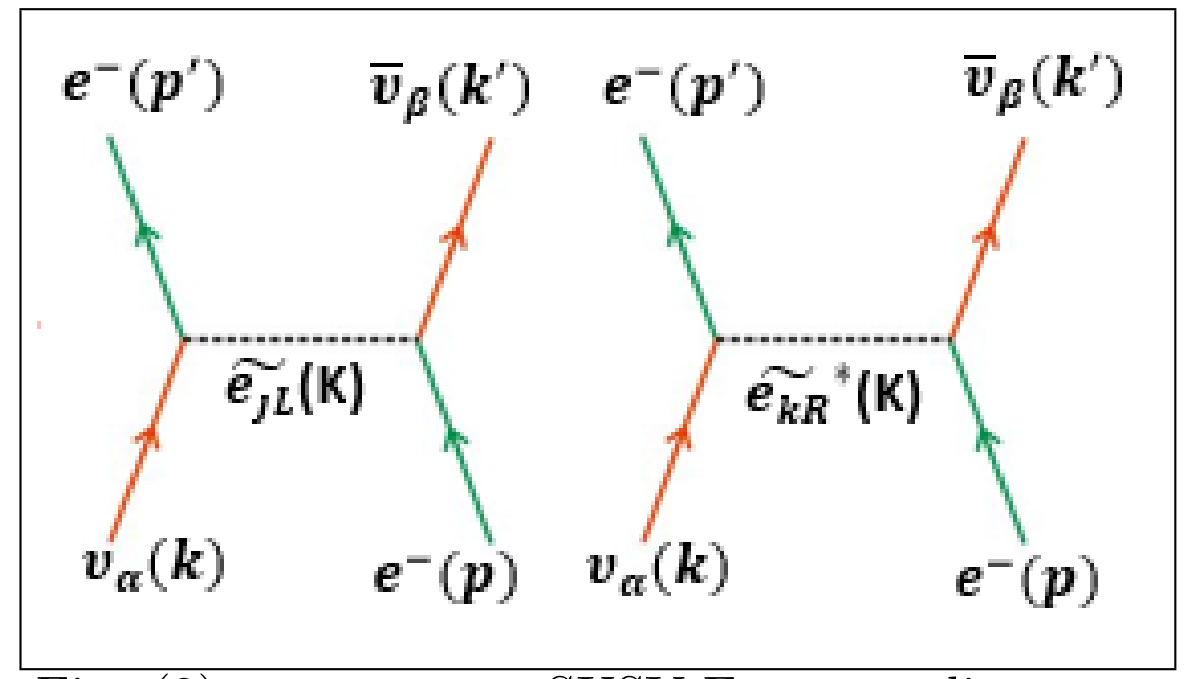

Fig. (3): $\nu_\alpha\, e \to \nu_\beta\, e$ SUSY Feynman diagrams

The relevant terms for process $\nu_\alpha\, e \to \nu_\beta\, e$ are $\tilde{e}_{jL}\, \bar{e}_{kR}\, \upsilon_{iL}$ and $\tilde{e}^*_{kR}\, (\bar{\upsilon}_{iL})^c e_{jL}$ can proceed through Feynman diagrams Fig.(3)

The R-parity Lagrangian for interaction of $\tilde{e}_{jL}\, \bar{e}_{kR}\, \upsilon_{iL}$ and $\tilde{e}^*_{kR}\, (\bar{\upsilon}_{iL})^c e_{jL}$ can be written in Eq.(15) and Eq.(16) respectively

$$L^a_{R_p} = -\frac{8G_F}{\sqrt{2}} A^{ee}_{\alpha\beta} \left[\bar{\nu}_\beta\, (k')\gamma_\mu\, P_L\, \nu_\alpha\, (k) \bar{e}(p')\gamma^\mu P_R\, e(p)\right] \tag{15}$$

$$L^b_{R_p} = -\frac{8G_F}{\sqrt{2}} B^{ee}_{\alpha\beta} \left[\bar{\nu}_\beta\, (k')\gamma_\mu\, P_L\, \nu_\alpha\, (k)\right] \left[\bar{e}(p')\gamma^\mu P_L\, e(p)\right] \tag{16}$$

where $A^{ee}_{\alpha\beta} \equiv \frac{\sqrt{2}}{8G_F} \sum_j \frac{\lambda_{\alpha j1}\, \lambda^*_{\beta j1}}{\tilde{m}^2_{jL}}$; $B^{ee}_{\alpha\beta} \equiv \frac{\sqrt{2}}{8G_F} \sum_k \frac{\lambda_{\beta 1k}\, \lambda^*_{\alpha 1k}}{\tilde{m}^2_{kR}}$.

$\tilde{m}_{jL}$ and $\tilde{m}_{kR}$ are masses of generation dependent exchanged s-leptons. By adding Eq.(15 and 16) we get R-parity effective Lagrangian

$$L_{int(R_p)} = -\frac{8G_F}{\sqrt{2}} \left[\left\{\bar{\nu}_\beta\, (k')\gamma_\mu\, P_L\, \nu_\alpha\, (k)\right\} \left\{\bar{e}(p')\gamma^\mu \left(A^{ee}_{\alpha\beta}\, P_R + B^{ee}_{\alpha\beta}\, P_L\right) e(p)\right\}\right] \tag{17}$$

| CL level | **Limits on SUSY parameters** $\sum_j \lvert\lambda_{1j1}\rvert$ | |
|---|---|---|
| | LSND | LAMPF E225 |
| 68% | $\sum_j \lvert\lambda_{1j1}\rvert \leq 0.26$ | $\sum_j \lvert\lambda_{1j1}\rvert \leq 0.27$ |
| 90% | $\sum_j \lvert\lambda_{1j1}\rvert \leq 0.30$ | $\sum_j \lvert\lambda_{1j1}\rvert \leq 0.31$ |
| 95% | $\sum_j \lvert\lambda_{1j1}\rvert \leq 0.31$ | $\sum_j \lvert\lambda_{1j1}\rvert \leq 0.32$ |

TABLE IV: R-parity violating coupling parameter limits at different CL levels

For $\nu_e\, e$ elastic scattering $B^{ee}_{\alpha\beta} \equiv \frac{\sqrt{2}}{8G_F}\sum_k \frac{\lambda_{\beta 1k}\,\lambda^*_{\alpha 1k}}{\tilde{m}^2_{kR}}$ becomes zero due to anti-symmetric nature of $\lambda_{11k}\,\lambda^*_{11k}$ . Total cross-section for $\nu_e\, e \to \nu_e\, e$ and $\nu_\mu\, e \to \nu_\mu\, e$ are

$$\sigma^{\nu_e} = 4\sigma_o \left[\left(1+\frac{g_L}{2}\right)^2 + \frac{1}{3}\left(\frac{g_R}{2} + \frac{1}{\sqrt{2}G_F}\sum_j \frac{|\lambda_{1j1}|^2}{\tilde{m}^2_{jL}}\right)^2\right] \tag{18}$$

$$\sigma^{\nu_\mu} = 4\sigma_o \left[\left(\frac{g_L}{2} + \frac{1}{\sqrt{2}G_F}\sum_k \frac{|\lambda_{21k}|^2}{\tilde{m}^2_{kR}}\right)^2 + \frac{1}{3}\left(\frac{g_R}{2} + \frac{1}{\sqrt{2}G_F}\sum_j \frac{|\lambda_{2j1}|^2}{\tilde{m}^2_{jL}}\right)^2\right] \tag{19}$$

At $\sum_k \tilde{m}_{kR} = \sum_j \tilde{m}_{jL} = 100GeV$ and using $\sigma^{\nu_e} = \sigma^{\nu_e}_{\text{exp}}$ from LSND and LAMPF E225 experiments, we plot Eq.(18) at 68%, 90% and 95% upper CL shown in Fig.(4).

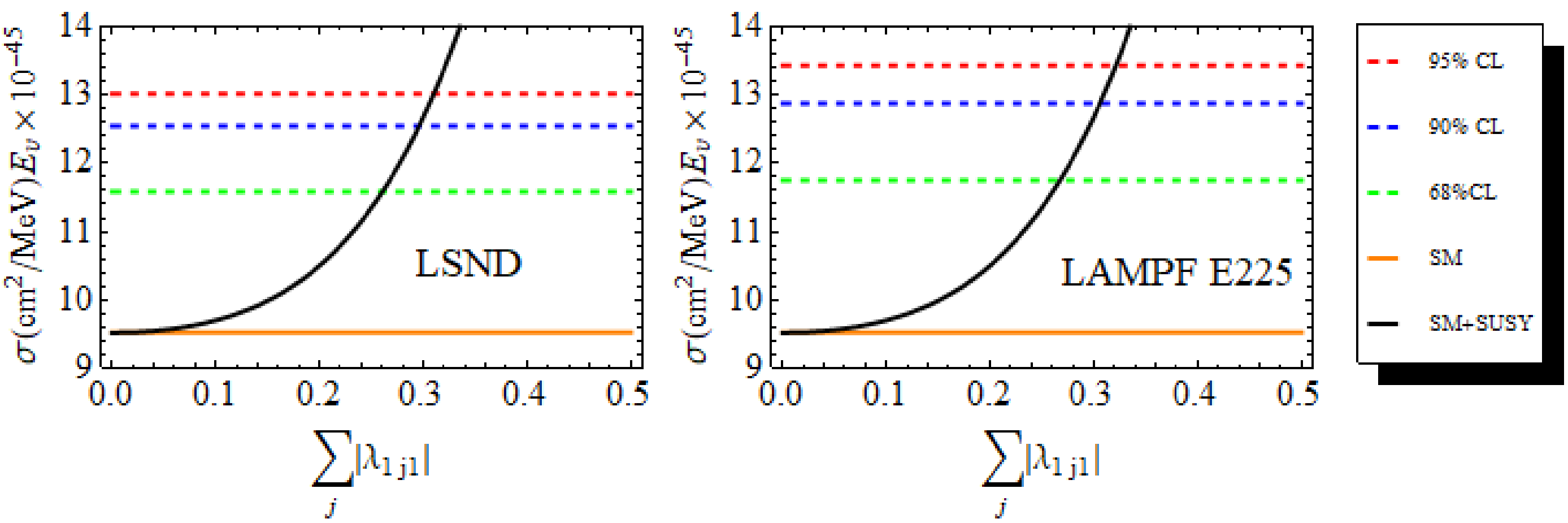

Fig. (4): Upper limit of R-parity violating SUSY parameters $\sum_j |\lambda_{1j1}|$ at different C.L

In the Fig.(4) it is clear that limits on $R\!\!\!/_p$ SUSY Model coupling parameter should lie in the interval mentioned in Table-IV. The SM cross-section can be obtain only by setting $\sum_j |\lambda_{1j1}| = 0$.

We analyze Eq.(19) by using the experimental values of CHARM-II, BNL-COL and BNL-E734 and obtained the Fig.(5) that reveal the sensitivity of $R\!\!\!/_p$ SUSY Model coupling parameters ($\sum_j |\lambda_{2j1}|$ and $\sum_k |\lambda_{21k}|$) to each other.

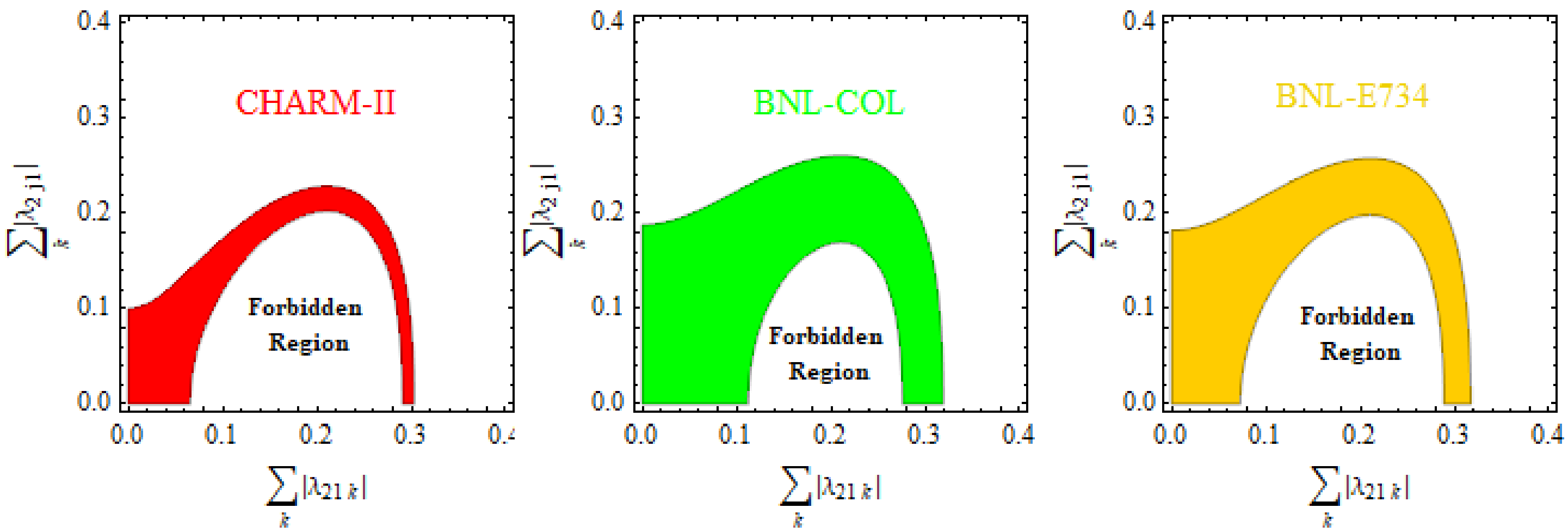

Fig. (5): Allowed and forbidden regions for R-parity violating SUSY coupling parameters $\sum_j |\lambda_{2j1}|$, $\sum_k |\lambda_{21k}|$ at 90%CL with $\sum_j \tilde{m}_{kR} = \sum_j \tilde{m}_{jL} = 100GeV$.

The overlap allowed and forbidden region wise sensitivity of $\sum_j |\lambda_{2j1}|$ and $\sum_k |\lambda_{21k}|$ for CHARM-II, BNL-COL and BNL-E734 can be seen in Fig. (6)

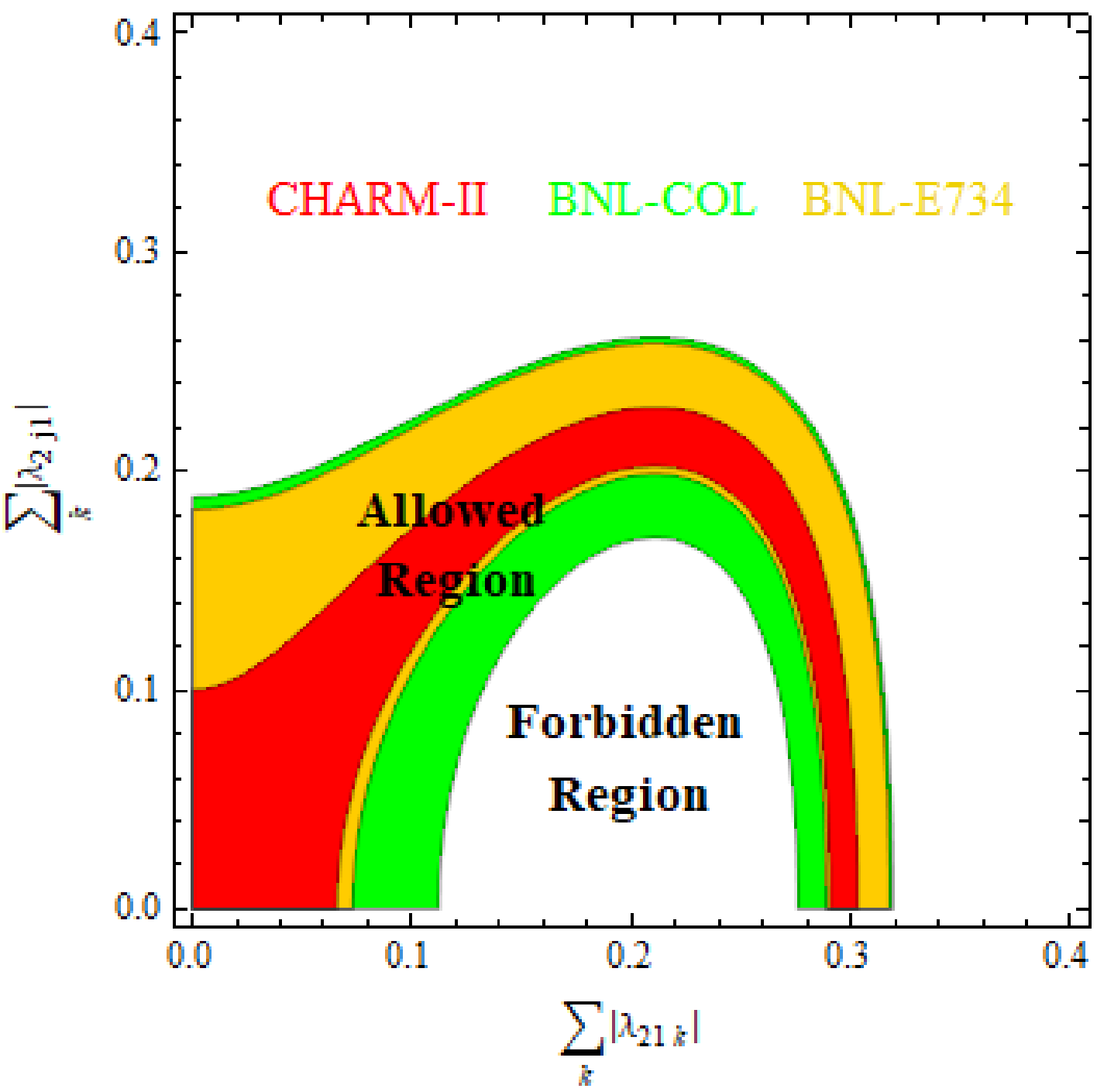

Fig. (6): Physical allowed and forbidden region for R-parity violating SUSY coupling parameters $\sum_j |\lambda_{2j1}|$ and $\sum_k |\lambda_{21k}|$ at 90%CL with $\sum_k \tilde{m}_{kR} = \sum_j \tilde{m}_{jL} = 100GeV$ from

By using single coupling dominances hypothesis, we find the limit on R-parity violating SUSY coupling parameters given in Table-V.

| **Limits on** $\sum_k \lvert\lambda_{21k}\rvert$ **and** $\sum_j \lvert\lambda_{2j1}\rvert$ | |
|---|---|
| CHRAM-II | |
| $\sum_k \lvert\lambda_{21k}\rvert \leq 0.07$ and | $\sum_j \lvert\lambda_{2j1}\rvert \leq 0.10$ |
| $0.29 \leq \sum_k \lvert\lambda_{21k}\rvert \leq 0.30$ | |
| BNL-E734 | |
| $\sum_k \lvert\lambda_{21k}\rvert \leq 0.11$and | $\sum_j \lvert\lambda_{2j1}\rvert \leq 0.19$ |
| $0.29 \leq \sum_k \lvert\lambda_{21k}\rvert \leq 0.32$ | |
| BNL-COL | |
| $\sum_k \lvert\lambda_{21k}\rvert \leq 0.11$ and | $\sum_j \lvert\lambda_{2j1}\rvert \leq 0.19$ |
| $0.29 \leq \sum_k \lvert\lambda_{21k}\rvert \leq 0.32$ | |

TABLE V: R-parity violating SUSY coupling parameter limits at 90 percent CL levels

By direct Comparison of Eqs.(9 and 17) we get the relation between MI coupling parameters and R-parity violating SUSY Model coupling parameters.

$$\varepsilon^{eL}_{\alpha\beta} = \frac{1}{\sqrt{2}G_F}\sum_k \frac{\lambda_{\beta 1k}\,\lambda^*_{\alpha 1k}}{\tilde{m}^2_{kR}} \tag{20}$$

$$\varepsilon^{eR}_{\alpha\beta} = \frac{1}{\sqrt{2}G_F}\sum_j \frac{\lambda_{\alpha j1}\,\lambda^*_{\beta j1}}{\tilde{m}^2_{jL}} \tag{21}$$

## 5. RESULTS AND DISCUSSION

We have performed a complete survey of $\nu_\alpha\, e \rightarrow \nu_\beta\, e$ scattering and calculate related percentage error. The major contribution in the cross-section of $\nu_e\, e$ and $\nu_\mu\, e$ is found to be coming from SM as mentioned in Table-I., shows the size of discrepancies between SM and experimental cross-section for the process $\nu_e\, e \rightarrow \nu_e\, e$ and $\nu_\mu\, e \rightarrow \nu_\mu\, e$ at 68%, 90% and 95% Confidence levels. These discrepancies give us a hint for the possibility of NSI and the size of discrepancies shows available room for NSI. Although the discrepancy between experimental value and SM prediction is very small but due to non-zero masses of neutrinos, we cannot neglect the NSI. The deviation from experimental limits places limits on new physics(NP) parameters at different error levels. By using advance neutrino experiments i.e. super-beams, $\beta-$beams or neutrino factory, we can correctly estimate uncertainties and discover NP within them.

In order to achieve this goal both Model Independent (MI) and Model Dependent (MD) ( R-parity violating SUSY Model), detailed analysis have been performed and relationship between MI and MD parameters is established.

In MI analysis the non-standard interaction is incorporated through non universal flavour diagonal couplings. The analysis is performed involving total cross-section for $\nu_e\, e \rightarrow \nu_e\, e$ and $\nu_\mu\, e \rightarrow \nu_\mu\, e$ as a sub-leading effect.

We discuss the sensitivity of flavour diagonal NSI Yukawa couplings parameters ($\varepsilon^{eL}_{ee}$ ,$\varepsilon^{eR}_{ee}$ and $\varepsilon^{eL}_{\mu\mu}$ ,$\varepsilon^{eR}_{\mu\mu}$ ) and identify the physically allowed and forbidden regions for these parameters under the experimental constraints of LSND, LAMPF-E225, CHARM-II, BNL-COL and BNL-E734 experimental shown in Figs.(8-13). We found limits on these couplings given in Table-II & III.

For MD (R-parity violating SUSY Model) analysis, the most general R-parity violating superpotential is introduced and R-parity violating currents pertaining to the process are also discussed. We have calculated elastic scattering cross-section $\sigma_{elastic}$ (SM+R-parity SUSY Model) for $\nu_e\, e \rightarrow \nu_e\, e$ and $\nu_\mu\, e \rightarrow \nu_\mu\, e$ in term of NUFD R-parity violating SUSY Model coupling parameters ($\sum_k |\lambda_{\alpha 1k}|$ , $\sum_j |\lambda_{\alpha j1}|$).

The contribution of $\sum_k |\lambda_{\alpha 1k}|$ becomes zero due to the anti-symmetry present in first two indices and only $\sum_j |\lambda_{\alpha j1}|$ contributes for $\nu_e\, e$ elastic scattering. We also plot $\sigma_{elastic}$ (SM+R-parity SUSY Model) vs $\sum_j |\lambda_{1j1}|$ for $\nu_e\, e \rightarrow \nu_e\, e$ and set the limits (given in Table-IV) by using LSND and LAMPF-E225 experimental value as shown in Fig.(4).

We indicate the allowed and forbidden physical regions for $\not{R}_p$ SUSY Model coupling parameters $\sum_j |\lambda_{2j1}|$and

$\sum_k |\lambda_{21k}|$ by using the experimental values of CHARM-II, BNL-COL and BNL-E734. Now the situation is quite different from MI scenario after incorporating SUSY propagating particles. The shapes of plots have changed and making the overall picture involving coupling parameters ($\sum_j |\lambda_{2j1}|, \sum_k |\lambda_{21k}|$) more clear as shown in Figs.(16 and 17). We have also calculated the limits on ($\sum_j |\lambda_{2j1}|, \sum_k |\lambda_{21k}|$) within allowed region (given in Table-V)

We have also developed the relation between MI coupling parameters $\left(\varepsilon_{ee}^{eL}, \varepsilon_{ee}^{eR}\right)$ and R-parity violating SUSY Model coupling parameters ($\sum_k |\lambda_{\alpha 1k}|, \sum_j |\lambda_{\alpha j1}|$) in term of Fermi coupling constant $G_F$ and mass of SUSY propagator Eqs.(20 and 21).

---

Our work for NSI similar to NSI tree level bounds of [1], but we use LSND, CHARM-II along with BNL-COL, BNL-E734 and LAMPF-E225 data instead of CHARM, NuTeV, KamLAND and SNO/SK data. Similarly authors of [2] used LSND data for neutrino and TEXONO antineutrino data for the specification of allowed region where one coupling hypothesis is not used but both coupling parameters are used, one as a function of other parameter is plotted. Our bounds are improved as compared to previously existing bounds For Supersymmeteric analysis we specify exact region for SUSY contribution for those process which were allowed at tree level processes in SM by using above mentioned experimental data, which is unique as compared to previous studies. We are also finding limits by using confidence levels 68% and 95% which was not done by the previous authors [1, 2]. We also develop the relationship between SUSY and NSI parameters.